\documentclass[aps,pra,preprint,showpacs,superscriptaddress,amsfonts,floatfix]{revtex4}
\usepackage{graphics}

\usepackage[dvips]{graphicx}
\usepackage{amsmath}

\begin{document}

%
%

\title{
Dynamic spectral aspects of interparticle correlation}

%
%
%
\author{I. Nagy}
\affiliation{Department of Theoretical Physics,
Institute of Physics, \\
Budapest University of Technology and Economics, \\ H-1521 Budapest, Hungary}
\affiliation{Donostia International Physics Center, P. Manuel de
Lardizabal 4, \\ E-20018 San Sebasti\'an, Spain}
\author{I. Aldazabal}
\affiliation{Centro de F{\'i}sica de Materiales (CSIC-UPV/EHU)-MPC,
P. Manuel de Lardizabal 5, \\ E-20018 San Sebasti\'an, Spain}
\affiliation{Donostia International Physics Center, P. Manuel de
Lardizabal 4, \\ E-20018 San Sebasti\'an, Spain}
\author{M. L. Glasser}
\affiliation{Department of Physics, Clarkson University, Potsdam,\\
New York 13699-5820, USA}
\affiliation{Donostia International
Physics Center, P. Manuel de Lardizabal 4, \\ E-20018 San
Sebasti\'an, Spain}

\date{\today}
\begin{abstract}

Time-dependent quantities are calculated in the linear response
limit for a correlated one dimensional model atom driven by an
external  quadrupolar time-dependent field. Besides the analysis of
the time-evolving energy change in the correlated two-particle
system, and orthogonality of initial and final states, Mehler's
formula is applied in order to derive a point-wise decomposition of
the time-dependent one-matrix in terms of time-dependent occupation
numbers and time-dependent orthonormal, natural orbitals. Based on
such exact spectral decomposition on the time domain, R\'enyi's
entropy is also investigated. Considering the structure of the exact
time-dependent one-matrix, an independent-particle model is defined
from it which contains exact information on the single-particle
probability density and probability current. The resulting
noninteracting auxiliary state is used to construct an effective
potential and discuss its applicability.

\end{abstract}


\maketitle

\section{Introduction and motivations}

There are quantum mechanical problems in which the Hamiltonian depends explicitly on time,
for example, the interaction of the system with an external
time-dependent field where $H(x,t)\, =\, H_0(x)+H'(x,t)$. In such cases, the system does not
remain in any stationary state, and the
behavior of it is governed by the time-dependent Schr\"odinger equation
\begin{equation}
i\, \hbar\, \frac{\partial\, \Psi(X,t)}{\partial\, t}\, =\, H(X,t)\, \Psi(X,t). \nonumber
\end{equation}
The problem will now be an initial-value (Cauchy) problem of solving
this partial differential equation for a prescribed initial
condition at $t_0$. When problems of this short are discussed
formally, it is common to speak of the perturbation ($H'$) as
causing {\it transitions} between eigenstates of $H_0$. If this
statement is interpreted to mean that the state has changed from its
inital value to a final value, than it is incorrect. The effect of a
time-dependent perturbation is to produce a nonstationary state,
rather than to cause a jump from one stationary state to another
which were determined by considering boundary conditions solely.

Experiments open the possibility to investigate dynamical properties
of confined systems. In that field, of particular interest is the
response of the system to time-dependent variations of the confining
field. Then, even assuming complete isolation  of the many-body
confined system from the environment, there is the question of how
the correlation properties change. Furthermore, one can consider an
atom in its ground state. If at times $t>t_0$ it becomes subject to
a time-varying potential energy caused by a charged heavy projectile
passing by, its electron cloud will be shaken up so that its energy
increases. The energy change, related to the stopping power, is a
measurable quantity. However, when we have shake-up processes, the
application of one-particle auxiliary pictures is not necessarily
useful {\it a priori}.

Motivated by such problems, in this work we consider a simplified
interacting model system introduced by Heisenberg
\cite{Heisenberg26,Moshinsky68,Ballentine98} in the early days of
quantum mechanics
\begin{equation}
\hat{H}_0(x_1,x_2)\, =\, -\, \frac{\hbar^2}{2m}\left(\frac{d^2}{dx_1^2}\, +
\frac{d^2}{dx_2^2}\right) +\frac{m}{2}\, \omega_0^2({x}_1^2+{x}_2^2)
-\frac{m}{2}\, \Lambda\, \omega_0^2({x}_1-{x}_2)^2,
\end{equation}
where $\Lambda\in{[0,0.5]}$ measures the strength of repulsive
interparticle interaction. We will use ideas and methods
\cite{Popov70,Kagan96,Campo13} for solving the quantum motion of a
particle in a harmonic oscillator with time-dependent frequency, by
adding a quadrupolar ($Q$) time-dependent perturbation, of
$\hat{H}'(t\rightarrow{-\infty})=\hat{H}'(t\rightarrow{\infty})=0$
character, to the above ground-state Hamiltonian
\begin{equation}
\hat{H}'(x_1,x_2,t)\, =\, \frac{m}{2}\, Q\, F(t)\, ({x}_1^2+{x}_2^2).
\end{equation}
In the field of heavy-particle interaction with atoms, such term could mimic the shake-up process
due to close encounters. There, as was stressed \cite{Fermi40} by Fermi,
quantum mechanics is needed since Bohr's classical treatment is
valid \cite{Bohr13,Dobson94} only for dipolar perturbation.

One might think that a two-particle model is a bit trivial to test
time-dependent many-body methods. But, this is not the case. Indeed,
it is the few-body correlated dynamics that serve \cite{Bauer13} as
benchmark for methods beyond, for instance, Time Dependent Density
Functional Theory, TD-DFT. Furthermore, to the theory of breathing
modes of many-body systems in harmonic confinement, it was utilized
\cite{Brabec13} to replace a given (say, with Coulombic
interparticle interaction) Hamiltonian with a quadratic Hamiltonian
for which an analytic solution exist. A mapping at the Hamiltonian
level could be a practical version of the isospectral deformation
discussed earlier \cite{Dreizler86} in the light of the,
unfortunately, formal character of pure existence theorems and
mapping lemmas behind TD-DFT.

We believe that experience about relevant details in interacting
systems is best governed by studying time-dependent quantities based
on exactly solvable instructive examples. In particular, the
simultaneous treatment of the notorious kinetic and interparticle
energy components could be more accurate than in TD-DFT where both
are approximated. Besides, our result allows a transparent
implementation of the mapping-formalism \cite{Ullrich12,Leeuwen15}
designed to construct independent-particle potentials (and to
discuss \cite{Schirmer07} an alarming paradox arising there) by
using exact probabilistic quantities of the correlated model as
inputs.

As a last motivation, we note that there have been efforts to
clarify the basic relations between interparticle correlation and
information-theoretic measures for entanglement. For instance, the
stationary ground-state of the correlated model applied in the
present work has already been investigated in this respect. A
surprising duality between R\'enyi's entropies characterizing
entangled systems, with attractive ($\Lambda<0$) and repulsive
interparticle interactions was found
\cite{Pipek09,Gross13,Glasser13} and explained \cite{Schilling14}
recently. The question of how such a remarkable duality will change
when the correlated system is perturbed by a time-dependent external
field is an exciting one of broad relevance. One of our goals here
is to provide an answer.

We stress that we restrict ourselves to the linear-response limit,
by taking our $QF(t)$ sufficiently small or sudden, since the main
goal is on the time-dependent spectral aspects of
correlation-dynamics and not on comparison with data. However, we
should note that similarly to the Born series of stationary
scattering theory, an order-by-order expansion could result in only
an asymptotic series. Indeed, higher-order response to external
field is an important topic in the energy change during shake-up
dynamics of a nucleus \cite{Robinson61} or an atom
\cite{Merzbacher74}. The idea that by renormalization of the kinetic
energy one could \cite{Prelovsek10} extend the validity of
linear-response formalism is also a challenging one. Such questions
need accurate solutions in linear and nonlinear responses. The
nonlinear version of the exact determination of energy changes in
our correlated model system will be published separately.

The rest of the paper is organized as follows. Section II contains
our theoretical results. Section III is devoted to a short summary
and few relevant comments. We will use natural, rather than atomic,
units in this work, except where the opposite is explicitly stated.

\section{Results and discussion}

By introducing the normal coordinates
$X_1\equiv{(x_1+x_2)/\sqrt{2}}$ and $X_2\equiv{(x_1-x_2)/\sqrt{2}}$,
one can easily rewrite \cite{Moshinsky68} the unperturbed
Hamiltonian in the form
\begin{equation}
\hat{H}_0(X_1,X_2)\, =\, -\, \frac{\hbar^2}{2m}\left(\frac{d^2}{dX_1^2}\, +
\frac{d^2}{dX_2^2}\right) +\frac{m}{2}\, \omega_1^2\, X_1^2 + \frac{m}{2}\, \omega_2^2\, X_2^2,
\end{equation}
where  $\omega_1\equiv{\omega_0}$ and
$\omega_2\equiv{\omega_0\sqrt{1-2\Lambda}}$ denote the resulting
independent-mode frequencies. It is this separated form which shows
that a time-dependent perturbation of dipolar character,
$D(t)(x_1+x_2)=[\sqrt{2}D(t)]\, X_1$, will couple only to one normal
mode characterized by the unperturbed angular frequency
$\omega_1=\omega_0$. Therefore, such perturbation
\cite{Bohr13,Robinson61,Merzbacher74,Dobson94} produces an energy
change independent of the correlated aspect of our model.

In the case investigated here, we have $QF(t)\,
(x_1^2+x_2^2)=QF(t)(X_1^2+X_2^2)$, i.e., there is time-dependent
quadrupolar perturbation in both independent normal modes, which
will evolve in time independently. Thus, we proceed by one
oscillator [$\hat{h}(X,t)$] in a time-dependent harmonic
confinement, following the established theoretical path, where one
has to solve a time-dependent Schr\"odinger equation, of the form
given by Eq. (1), with
\begin{equation}
\hat{h}(X,t)\, =\, -\, \frac{\hbar^2}{2m}\, \frac{d^2}{dX^2}\, + \frac{m}{2}\, \Omega^2(t)\, X^2 .
\end{equation}
The solution rests on making proper changes of the time and distance
scales \cite{Popov70,Kagan96} in order to consider frequency
variations in $\Omega(t)$ as it changes from $\Omega_0$ during
time-evolution. The exact mode, a nonstationary evolving state
$\psi[X,\Omega_0,R(\Omega_0),t]$, contains these scales as
\begin{equation}
\psi(X,\Omega_0,R,t)=\left[\frac{m\, \Omega_0}{\hbar\, R^2(t)\, \pi}\right]^{1/4}\, \exp\left[-\frac{X^2}{2}\, \frac{m\, \Omega_0}{\hbar \, R^2(t)}
\left(1-i\, \frac{R(t)\dot{R}(t)}{\Omega_0}\right)\right]\, e^{-i\, \gamma(t)}.
\end{equation}

This exact quantum mechanical, time-dependent solution is obtained \cite{Popov70,Kagan96} by
considering a classical equation of motion
\begin{equation}
\ddot{\xi}(t) +\, \Omega(t)\, \xi(t)\, =\, 0,
\end{equation}
with complex $\xi(t)=R(t)\exp[i \gamma(t)]$ substitution, where the real $R(t)$ gives the length scale at the instant $t$.
The nonlinear differential equation, determining this scale becomes
\begin{equation}
\ddot{R} + R(t)\, \Omega^2(t)\, =\, \frac{\Omega_0^2}{[R(t)]^3},
\end{equation}
after taking $\dot{\gamma}(t)\equiv{\Omega_0/R^2(t)}$. The initial
conditions are $R(t_0=0)=1$ and $\dot{R}(t=0)=0$. Since we are
dealing with a weak external perturbation in this work on spectral
dynamics, we solve Eq. (7) via the substitutions $R(t)=1+r(t)$ and
$\dot{R}(t)=\dot{r}(t)$.


\noindent With $r\ll{1}$ to Eq. (7),  we get a
forced-oscillator-like differential equation
\begin{equation}
\ddot{r}(t)\, +\, (2\Omega_0)^2\, r(t)\, =\, Q\, F(t),
\end{equation}
which is treated by going to the complex notation
$w(t)=[\dot{r}(t)+i(2\Omega_0)r(t)]$ in order to derive a
first-order differential equation for $w(t)$. Remarkably, Eq. (8)
shows transparently that the breathing frequency \cite{Brabec13} is
precisely $2\times{\Omega_0}$, i.e., the double of the confinement
frequency characterizing the ground-state mode of the unperturbed
$\hat{h}_0(X)$.

Once an explicit form for  $F(t)$ is prescribed to be used in both
($i=1,2$) differential equations for $r_i(t)$ [considering the two
independent {\it modes}, with $\Omega_0\rightarrow{\omega_1}$ and
$\Omega_0\rightarrow{\omega_2}$] and the solutions for
$r_i(\omega_i,t)$ are found, the time-dependent wave function
becomes
\begin{equation}
\Psi(X_1,X_2,t)\, =\, \psi(X_1,\omega_1,R_1,t)\, \psi(X_2,\omega_2,R_2,t),
\end{equation}
which is valid for $|Q|<{\omega_i}$, and to which the form for
$\psi(X_i)$ is given by Eq. (5). By using this
normal-coordinate-based representation for the exact wave-function
we can easily calculate with it and the {\it unperturbed}
Hamiltonian, the expectation values of the kinetic and potential
energy components, $<K(t)>$ and $<V(t)>$, respectively. We get for
these quantities
\begin{equation}
<K(t)>\, \, =\, \, \sum_{i=1}^2\, \frac{\hbar\, \omega_i}{4}\,
\left[\frac{1}{R_i^2(\omega_i,t)}\, +
\frac{\dot{R}_i^2(\omega_i,t)}{\omega_i^2} \right]
\end{equation}
\begin{equation}
<V(t)>\, \, =\, \sum_{i=1}^2\, \frac{\hbar\, \omega_i}{4}\,
R^2_i(\omega_i,t).
\end{equation}

In the linear-response limit, the total energy {\it change} $\Delta
E(t)\equiv{E(t)-E(t_0)}$ becomes
\begin{equation}
\Delta E(t)\, =\,  \sum_{i=1}^2\, \Delta E_i(t)\,\Rightarrow\,{
\sum_{i=1}^2\, \frac{\hbar\, \omega_i}{4}\,
\left[\frac{(2\omega_i)^2\, r_i^2(\omega_i,t)+
\dot{r}_i^2(\omega_i,t)}{\omega_i^2}\right]}.
\end{equation}
To arrive at the consistent r.h.s. above, which is valid for weak
external fields, we linearized Eqs. (10-11). In this case [with Eq.
(8)] one gets $\dot{E}_i(t)=(\hbar\omega_i/4)[2\,
\dot{r}_i(\omega_i,t)\, Q \, F(t)/\omega_i^2]$ for the rate. Next,
we derive for the overlap $O(t)=|\Psi(t_0)\Psi(t)|^2$ the following
expression
\begin{equation}
O(t)\, =\,  \frac{1}{\sqrt{1+\Delta E_1(t)/\hbar \omega_1}\, \,
\sqrt{1+\Delta E_2(t)/\hbar \omega_2}},
\end{equation}
in terms of $\Delta E_i(t)$. This quantity is a measure of
correlated (when $\omega_1\neq{\omega_2})$ dynamics. Its product
form reflects the independence of two normal modes during
propagation. When we are close to the stability limit at
$\Lambda\rightarrow{0.5}$, even a weak external field could result
in an almost perfect orthogonality at certain, $F(t)$-dependent,
time $t>t_0$.

Now, we turn to quantities which show explicitly the entangled
nature of the correlated two-particle system in the time domain. By
rewriting the wave function in terms of original coordinates, we
determine the reduced single-particle density matrix
[$\Gamma_1(x_1,x_2,t)$] from
\begin{equation}
\Gamma_1(x_1,x_2,t)\, =\, \int_{-\infty}^{\infty}\, dx_3\,
\Psi^{*}(x_1,x_3,t)\, \Psi(x_2,x_3,t),
\end{equation}
After a long, but straightforward, calculation we obtain
\begin{equation}
\Gamma_1(x_1,x_2,t)=\phi_s(x_1,t)\, \phi_s^{*}(x_2,t)\,
e^{-\frac{m}{2\hbar}A(t)(x_1-x_2)^2},
\end{equation}
where, for further clarifications below, we introduced the following
abbreviations
\begin{equation}
\phi_s(x,t)=\left[\frac{m \omega_s(t)}{\hbar \pi}\right]^{1/4}\,
e^{-\frac{m}{2\hbar}\omega_s(t)\, x^2[1-i\, \alpha(t)/\omega_s(t)]},
\end{equation}
\begin{equation}
\frac{\alpha(t)}{\omega_s(t)}\, =\, \frac{1}{2}\,
\left[\frac{R_1(\omega_1,t)\dot{R}_1(\omega_1,t)}{\omega_1}+\frac{R_2(\omega_2,t)\dot{R}_2(\omega_2,t)}{\omega_2}
\right]
\end{equation}
\begin{equation}
A(t)\, =\, \frac{1}{4}\, \frac{[\omega_1(t)-\omega_2(t)]^2
+[\dot{R}_1(\omega_1,t)/R_1(\omega_1,t) -
\dot{R}_2(\omega_2,t)/R_2(\omega_2,t)]^2} {\omega_1(t)+\omega_2(t)}
\end{equation}
with
$\omega_s(t)=2\omega_1(t)\omega_2(t)/[\omega_1(t)+\omega_2(t)]$,
where $\omega_i(t)=\omega_i/[R_i(\omega_i,t)]^2$. The one-matrix is
Hermitian, as it must be, since
$\Gamma_1(x_1,x_2,t)=\Gamma_1^{*}(x_2,x_1,t)$. Later we will give,
as in the stationary case \cite{Glasser13} earlier, a direct
spectral decomposition of this important Hermitian matrix in
Eq.(15), without considering, as more usual, an eigenvalue problem.

\newpage

The diagonal of the reduced single-particle density matrix  gives
the basic variable of TD-DFT, i.e., the one-particle probability
density
\begin{equation}
n(x,t)=\left[\frac{m \omega_s(t)}{\hbar \pi}\right]^{1/2}\,
e^{-\frac{m}{\hbar}\omega_s(t)\, x^2}.
\end{equation}
The single-particle probability current, $j(x,t)$, is calculated as
usual \cite{Ballentine98,Lein05,Nagy13} from
\begin{equation}
j(x,t)\, =\, \frac{2 \hbar}{m}\, Re\, \int\,
\Psi^{*}(x,x',t)\frac{\partial}{i\, \partial x}\, \Psi(x,x',t)\,
dx'.
\end{equation}
which, in terms of the functions introduced above, becomes
\begin{equation}
j(x,t)\, =\, \frac{2 \hbar}{m}\, x\, n(x,t)\, \alpha(t).
\end{equation}
These probability densities satisfy the continuity equation:
$\partial_t n(x,t)+\partial_x j(x,t)=0$.

By taking the $A(t)\rightarrow{0}$ substitution in Eq.(15), i.e.,
neglecting the important role of the interparticle coordinate, we
can define an auxiliary independent-particle model, where the
particles move in a {\it certain} external field. The wave function
of this modeling becomes
\begin{equation}
\Psi_s(x_1,x_2,t)\, =\, \phi_s(x_1,t)\, \phi_s(x_2,t).
\end{equation}
This approximate state will still result in the exact probability
density and exact probability current, by construction. So, one may
consider it as an {\it output} of TD-DFT. Indeed, using Eqs.
(6.50-6.52) of \cite{Ullrich12} or Eqs. (21-22) of \cite{Campo13},
an effective potential $V_{s}(x,t)$ is given by
\begin{equation}
V_{s}[x,t,\omega_s(t)]\, =\, \frac{1}{2}\omega_{s}^2(t)\, x^2\, -\,
\frac{1}{2}\left[\sqrt{\omega_{s}(t)}\, \frac{d^2}{d
t^2}\left(\frac{1} {\sqrt{\omega_{s}(t)}}\right)\right] x^2,
\end{equation}
at least upto a time-dependent constant \cite{Lein05,Nagy13}. In the
so-called adiabatic treatment in TD-DFT, one uses \cite{Ullrich12}
only the first term as a time-dependent external field.

But, with such a single-particle potential, which produces from a
time-dependent wave equation the $\Psi_s(x_1,x_2,t)$ state, we also
arrive at a dilemma. We know the exact Schr\"odinger Hamiltonian.
Especially, we know how the unperturbed Hamiltonian is modified by
adding a time-dependent quadrupolar perturbation to it. Furthermore,
since the perturbation acts only over a limited time in our problem,
in the Schr\"odinger picture we can follow the standard quantum
mechanical prescription to determine the energy change by
calculating expectation values with the exact wave function
$\Psi(X_1,X_2,t)$ and unperturbed Hamiltonian $\hat{H}_0(X_1,X_2)$.
In the light of $V_{s}[x,t,\omega_s(t)]$, the situation in the
orbital version of TD-DFT is not so clear. For instance, it will
contain a $\ddot{\omega}_s(t)$-proportional higher-order derivative
\cite{Ullrich12,Schirmer07,Campo13} . Quite unfortunately, in
TD-DFT, which is based \cite{Ullrich12,Leeuwen15,Schirmer07} on a
density-potential mapping lemma, we do not \cite{Dreizler86} know
universal functionals to determine an energy change via them.

\newpage

After the above remarks on open questions in applied DT-DFT, and
motivated partly by Dreizler's early suggestion \cite{Dreizler86} on
using an isospectral deformation ($d$) and a constrained search, we
turn to the decomposition of our reduced one-particle density
matrix, and, in addition, put forward an idea on the possibility of
using an {\it approximate} nonidempotent one-matrix,
$\Gamma_1^s(x_1,x_2,t)$, designed below. Following our experience
\cite{Glasser13} on the decomposition of the stationary
$\Gamma_1(x_1,x_2)$, we will use Mehler's formula
\cite{Erdelyi53,Koscik15} now to $\Gamma_1(x_1,x_2,t)$.

Thus, via the
$[\omega_s(t)+A(t)]=\bar{\omega}(t)[1+z^2(t)]/[1-z^2(t)]$ and
$A(t)=\bar{\omega}(t)2z(t)/[1-z^2(t)]$ correspondences, we introduce
two new variable $\bar{\omega}(t)$ and $z(t)$. The solutions are
\begin{equation}
\bar{\omega}(t)\, =\, \omega_s(t)\, \frac{1+z(t)}{1-z(t)},
\end{equation}
\begin{equation}
z(t)\, =\, \left[\frac{\omega_s(t)}{A(t)} + 1\right] -
\sqrt{\left[\frac{\omega_s(t)}{A(t)}+1\right]^2-1}.
\end{equation}
Now, in term of these we get our point-wise decomposition directly
\begin{equation}
\Gamma_1(x_1,x_2,t)\, =\, \sum_{l=0}^{\infty}\, [1-z(t)][z(t)]^l\,
\phi_l(x_1,t)\, \phi_l^{*}(x_2,t),
\end{equation}
where the $\phi_l(x,t)$ elements of the orthonormal basis set
(natural orbitals) are
\begin{equation}
\phi_l(x,t)\, =\,
\left[\frac{m\bar{\omega}(t)}{\hbar\pi}\right]^{1/4}\, e^{i\,
\frac{m}{2\hbar}\alpha(t)x^2}\, \frac{1}{\sqrt{2^l\, l!}}\,
e^{-\frac{m}{2\hbar}\bar{\omega}(t)x^2}\,
H_m(\sqrt{m\bar{\omega}(t)/\hbar}x).
\end{equation}

Of course, extracted quantities, like the exact one-matrix
$\Gamma_1(x_1,x_2,t)$ and the associated exact probability density
$n(x,t)$, contain less and less information then the wave function
$\Psi(x_1,x_2,t)$ and the pair-function
$n_2(x_1,x_2,t)=\Psi(x_1,x_2,t)\Psi^{*}(x_1,x_2,t)$. However, even
by these extracted quantities one can calculate the exact kinetic,
and the external-potential energies, respectively, while in TD-DFT
only the last energy is, in principle, exact. Thus, to formulate an
idea, which is based on inversion from {\it known} probability
density $n(x,t)$ and probability current $j(x,t)$ (i.e., from the
crucial \cite{Schirmer07,Rapp14} $\alpha(t)$ phase), we make a
rewriting
\begin{equation}
n(x,t)=\sum_{l=0}^{\infty}\, [1-z(t)][z(t)]^l
|\phi_l[x,t,\bar{\omega}(t)|]^{2}= \sum_{l=0}^{\infty}
[1-z_d(t)][z_d(t)]^l|\phi_l[x,t,\omega_d(t)]|^{2},
\end{equation}
which is valid if and only if $\omega_d(t)\, =\, \omega_s(t)\,
[1+z_d(t)]/[1-z_d(t)]$.

So, with one-particle probabilistic functions as input, we can
prescribe, from the r.h.s. of the above equation, a trace-conserving
$\Gamma_1^s(x_1,x_2,t)$ which has the form of $\Gamma_1(x_1,x_2,t)$.
Such nonidempotent form contains two (interconnected) parameters
$\omega_d(t)$ and $z_d(t)$, instead of the uniquely derived
$\bar{\omega}(t)$ and $z(t)$. Whether this deformed one-matrix could
result in, because of its {\it tunable} flexibility, an essentially
better approximation for the kinetic energy than the TD-DFT
[$z(t)=z_d(t)\equiv{0}$] where the one-matrix is idempotent,
requires future investigations within the framework of a constrained
search \cite{Dreizler86,Riveros14}.


The information-theoretic aspects of iterparticle interaction, i.e.,
of a crucial term in the Hamiltonian, may be considered via entropic
measures which do not use expectation values with physical
dimensions but use pure probabilities encoded in the occupation
numbers
\begin{equation}
P_l(t)\, \equiv{[1 - z(t)]\, [z(t)]^l},
\end{equation}
which are, in the present case, time-dependent quantities as well.
First, we calculate R\'enyi's entropy \cite{Renyi70}, with
$q\neq{1}$ and $q>0$, from
\begin{equation}
S_R(q,\Lambda,t)\, \equiv{\frac{1}{1-q}\,
\ln\frac{[1-z(t)]^q}{1-[z(t)]^q}}.
\end{equation}

The von Neumann entropy is obtained as a limiting case when
$q\rightarrow{1}$. It is given by
\begin{equation}
S_N(\Lambda,t)\, \equiv{- \ln[1-z(t)]\, -\frac{z(t)}{1-z(t)}\,
\ln[z(t)]}.
\end{equation}
In the knowledge of $z(t)$, these entropies characterize the
deviation from the stationary case described by $z(t_0)$, i.e.,
without external perturbation on the correlated two-body system. By
considering repulsive [$\Lambda\in(0,0.5)$] and attractive
[$\Lambda<0$] versions, we can investigate the duality
\cite{Glasser13,Schilling14} problem with respect to entropies, now
in the time domain. Notice, that the knowledge of R\'enyi entropies
for arbitrary values of $q$ provides, mainly from
information-theoretic point of view, more information than the von
Neumann entropy alone, since one could extract \cite{Calabrese15}
from them the full spectrum of the one-matrix. But, as Eq. (26)
shows transparently, to physical expectation values one also needs
the proper basis set.

Armed the above, strongly interrelated, theoretical details on the
dynamics of interparticle correlation we come to their
implementation. We use a simple model for illustration, with
$H'(t)\, =\, Q\, F(t) = Q\, \Theta(t)\exp(-\beta\, t)$. By tuning
the value of the effective rate of change (denoted by $\beta$) we
can easily investigate the sudden and adiabatic limits. We get
\begin{equation}
R_i(\Lambda,\omega_i,t)\, =\, 1 +\, \frac{Q}
{\beta^2+(2\omega_i)^2}\left[e^{-\beta t} -\cos(2\omega_i t) +
\frac{\beta}{2\omega_i}\sin(2\omega_i t)\right]
\end{equation}
\begin{equation}
\dot{R}_i(\Lambda,\omega_i,t)\, =\, \frac{Q}
{\beta^2+(2\omega_i)^2}\left[-\beta\, e^{-\beta t}
+(2\omega_i)\sin(2\omega_i t) + \beta\, \cos(2\omega_i t)\right]
\end{equation}
\begin{equation}
\Delta E_i(\Lambda,\omega_i,t)\, =\, \frac{\hbar\, \omega_i}{4}\,
\frac{Q^2}{\beta^2+(2\omega_i)^2}\,
\frac{1}{\omega_i^2}\, \left[1 + e^{-2\beta\, t} - 2\, e^{-\beta\,
t}\, \cos(2\omega_i t)\right]
\end{equation}
The sum ($i=1,2$) of $\Delta E_i(\Lambda,\omega_i,t)$ measures the
total deviation from the unperturbed ground-state energy
$E(t=0)=(\hbar/2)(\omega_1+\omega_2)$. In the kick-limit, where
$\beta t\rightarrow{0}$, we have $\Delta
E_i(\Lambda,\omega_i,t)\propto{\hbar \omega_i(Q\beta
t/\omega_i)^2/[\beta^2+(2\omega_i)^2]}$. As expected on physical
grounds, the change at long time, i.e., taking
$t\rightarrow{\infty}$ first, becomes a constant which tends to zero
in the $\beta\rightarrow{\infty}$ sudden limit. In the long-time
limit the overlap parameter of Eq. (13), which measures the
time-evolution of orthogonality between exact states, also tends to
a constant value.

Simple independent-particle  modelings of the correlated model
system could be based on effective ($e$), prefixed external fields
$(m/2)\omega^2_e(x^2_1+x^2_2)$ before time-dependent perturbation.
In such cases one gets $\Delta
E_e(t=\infty,\omega_e)=2\times{{(\hbar
\omega_e/4)(Q/\omega_e)^2/[\beta^2+(2\omega_e)^2]}}$ for the total
energy change, following the standard quantum mechanical averaging
procedure discussed at Eq. (23). Two options are analyzed here, and
a remarkable result is found, as Figure 1 signals. Namely, with
$\omega_e\rightarrow{\omega_s(t=0)}$, i.e., with the density-optimal
modeling, the corresponding energy {\it change} tends to the exact
one at high enough values of $\beta$, since $2 \omega_s^{-1}(t=0) =
(\omega_1^{-1}+\omega_2^{-1})$. Somewhat surprisingly, by using the
so-called energy-optimal Hartree-Fock \cite{Moshinsky68,Pipek09}
modeling, where $\omega_e\rightarrow{\omega_0\sqrt{1-\Lambda}}$, one
cannot get agreement with the exact result at any value of $\beta$.
For illustration, a convenient ratio-function is defined as
\begin{equation}
\mathcal{R}(\Lambda,\beta,t)\, =\, \frac{\Delta
E_e(\Lambda,\beta,t)}{\Delta E(\Lambda,\beta,t)},
\end{equation}
and plotted in Figure 1 as a function of the $\beta$ with fixed,
density-optimal, $\omega_e=\omega_s(t=0)$. We have checked, using
both sides of Eq. (12), that with a $Q=0.2$ value we are in the
linear-response limit, where the energy changes are proportional to
$Q^2$. Notice, that we use standard Hartree atomic units, where
$m=\hbar=1$, to both Figures of this work.

\begin{figure}
\scalebox{0.3}[0.3] {\includegraphics{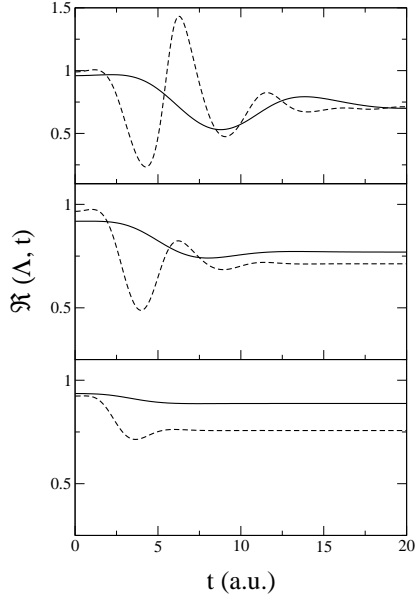}}
\caption {Ratio of the energy changes $\mathcal{R}(\Lambda,t)$, as a
function of $t$ which is measured in atomic units. The solid and
dashed curves refer to interparticle repulsion ($\Lambda=0.4$) and
attraction ($\Lambda=-2.0$), respectively. We used $Q=0.2$ and
$\omega_0=1$. The top, middle, and bottom panels refer,
respectively, to the $\beta=0.25$, $\beta=0.5$, and $\beta=1.0$
values. \label{figure1}}
\end{figure}

As we can see from the Figure, far from the sudden limit, i.e., at a
small $\beta$ value, the density-optimal modeling underestimates the
true energy change in the correlated case. We speculate that such
sensitivity close to the adiabatic limit could be, at least
partially, behind yet unclarified controversies in the applied
TD-DFT \cite{Zeb13,Mao14} to energy transfer by slow ions in
wide-gap insulators, where $\Lambda$ is positive of course. In
Section III, we will return to the informations given in Figure 1,
considering there the energy changes for negative $\Lambda$.


At the end of this Section, we turn to illustrations of the spectral
aspect of dynamic interparticle correlation. We apply von Neumann
entropy to get a precise insight into the time-dependence of an
information-theoretic measure. In the light of facts on
Hamiltonian-based measures, the energy-change and the wave-function
overlap at $t\rightarrow{\infty}$, such analysis is desirable. Based
on it, one can discuss, and maybe understood, challenging
interrelations between measures which are based on different
emphasizes within quantum mechanics. We stress here that the exact
energy-changes depend, as expected, on the sign of $\Lambda$.


%
\begin{figure}
\scalebox{0.3}[0.3] {\includegraphics{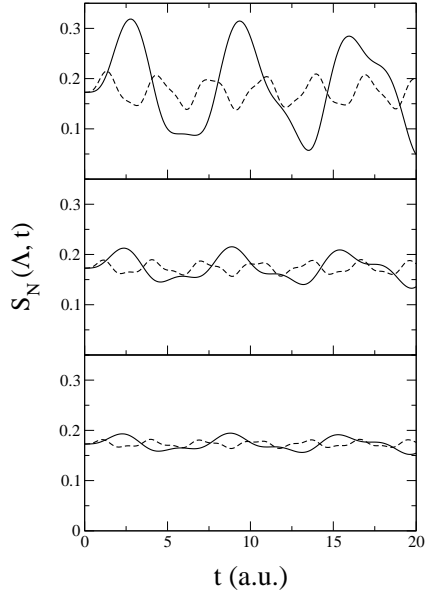}} \caption
{Information-theoretic, von Neumann, entropies $S_N(\Lambda,t)$, as
a function of the time $t$, which is measured in atomic units. The
solid and dashed curves refer to a moderate interparticle repulsion
$\Lambda=0.3$, and the corresponding (see the text) interparticle
attraction $\Lambda=-0.75$, respectively. Three values for the rate
parameter $\beta$, measured in inverse time-units, are employed to
illustration. Top panel: $\beta=2$. Middle panel: $\beta=5$. Bottom
panel: $\beta=10$. \label{figure2}}
\end{figure}

In Figure 2 we exhibit $S_N(\Lambda,t)$, by taking $\omega_0=1$,
$Q=0.2$, $\Lambda=0.40$ and $\Lambda=-2.0$. These couplings for
repulsion and attraction, respectively, are fixed to yield equal
entropies at $t=0$, in order to shed further light on duality
\cite{Pipek09,Glasser13,Schilling14}, now in the time-domain as
well. Notice, that in the unperturbed case to {\it any} allowed
[$\Lambda\in(0,0.5)]$ repulsive coupling there exists a
corresponding attractive ($\Lambda<0$) one for which the calculated
entropies are equal. For instance, when we are close to the
stability limit in the repulsive ($r$) case, i.e., when
$\Lambda_r\rightarrow{0.5}$, the corresponding attractive ($a$)
coupling becomes $\Lambda_a\rightarrow{(1/2)(2\Lambda_r -1)^{-1}}$.


The solid and dashed curves refer to repulsion and attraction, as we
mentioned. They are plotted as a function of the time measured in
atomic units. The rate-parameter of the time-dependent external
perturbation is described by  $\beta$ and measured in inverse units
of time. In order to discuss tendencies, we used three values for
this parameter. As expected, at very short times the entropies,
similarly to the energy changes, grow in time. After about 1-2
atomic units (i.e., about 25-50 attoseconds) they start to reflect
the oscillating character builded into the time-dependent occupation
numbers $P_l(t)$, the primary root of which is in the time-evolving
correlated wave function. The entropies will keep, in absence of an
environment-made dissipative \cite{Albrecht75,Tokatly13} coupling,
such oscillating behavior without limitation. From this point of
view, they signal the time-evolvement of the closed interacting
system. More importantly, one can see from the illustration that the
initial dual character of entropies, fixed at $t=0$, will {\it
disappear} due to a new physical variable, i.e., the time. Finally,
there are reductions in the oscillation-amplitudes by increasing
$\beta$. Clearly, at the so-called sudden limit
($\beta\rightarrow{\infty}$) for a time-dependent perturbation, they
will diminish as expected, since the system's reaction-rapidity is
limited by its normal mode frequencies.


In order to justify the above argument more directly, we take now a
somewhat less realistic modeling by using an abrupt quench of {\it
all} interactions at $t_0=0$ in the Hamiltonian.  In that case,
instead of Eqs. (33-34), one gets \cite{Kagan96} the following exact
expressions
\begin{equation}
R_i(\Lambda,\omega_i,t)\, =\, \sqrt{1+t^2\, \omega_i^2}
\end{equation}
\begin{equation}
\dot{R}_i(\Lambda,\omega_i,t)\, =\, \frac{t\,
\omega_i^2}{\sqrt{1+t^2\, \omega_i^2}}.
\end{equation}
There is no change in the system's kinetic energy since
$[1/R_i(\omega_i,t)]^2+[\dot{R}_i(\omega_i,t)/\omega_i]^2=1$. By
using time-dependent $A(t)$ and $\omega_s(t)$, which are needed to
Eq. (25), we get $z(t)=z(t=0)$ for all $t\geq{0}$ since
$[\omega_s(t)/A(t)]\equiv{2[2\sqrt{\omega_1\omega_2}/(\omega_1-\omega_2)]^2}$.
Thus, there is no entropy-{\it change} in the total-quench case.
This $S(t)=constant$ character could be a useful constraint in
practical attempts, with a good starting $n(x,t)$, using the
interconnected $z_d(t)$ and $\omega_d(t)$ variables of Eq. (28). One
can take $\omega_d(t)\Rightarrow{\omega_d/R^2(\omega_d,t)}$, where
$R(\omega_d,t)=\sqrt{1+t^2\omega_d^2}$.



\section{summary and comments}

Based on the time-dependent Schr\"odinger equation, an exact
calculation is performed for the time-evolving energy change, and
orthogonality of initial and final states, in a  correlated
two-particle model system driven by a weak external field of
quadrupolar character.  Besides, considering the structure of the
exact time-dependent one-matrix, an independent-particle model is
defined from it which contains exact information on the
single-particle probability density and probability current. The
resulting noninteracting auxiliary state is used to construct an
effective potential and discuss its applicability. We analyzed the
energy changes in a comparative manner, and pointed out (see, below
as well) the limited capability of an optimized independent-particle
modeling for the underlying time-dependent process.

A point-wise decomposition of the reduced one-particle density
matrix is given in terms of time-dependent occupation numbers and
time-dependent orthonormal orbitals. Based on such exact spectral
decomposition on the time domain, an entropic measure of inseparable
correlation is also investigated. It is found that the duality
behavior characterizing the stationary ground-state may disappear
depending on the rapidity of the external time-dependent
perturbation. At realistic rapidity, the undamped time-dependent
oscillations, found in an entropic measure, reflect the fact that
the correlated system evolves in time. In this respect, we have a
signal on the excited state, similarly as one can see in the
time-evolving probability density and probability current.


Our first comment is based on results obtained for the energy
changes, and plotted in Figure 1, by considering the sign of
$\Lambda$. In the case of an attractive interparticle interaction,
the case of a nucleus \cite{Moshinsky68}, the exact result for
$\Delta E(\Lambda<0,\beta,t)$ becomes bigger at moderate $\beta$
values than the result obtained within the density-optimal framework
$\Delta E_e(\Lambda<0,\beta,t)$. So, we speculate that a recent
result \cite{Stetcu15} on the excitation of a deformed nucleus
within TD-DFT could have the same, underestimating character in its
quadrupole channel, as the atomistic case has. Thus, the true
enhancement in excitations over the result based on the modeling of
Teller \cite{Teller48}, might also need further theoretical
refinement for that channel.


In the second comment, we focus on the role of the switching rate
$\beta$ in the case of energy changes. As a preliminary step to
future realistic calculations, here we would like to estimate a
characteristic interaction-time ($T\propto{\beta^{-1}}$) in
classical atom-atom collision. To do that, we model the
three-dimensional screened interaction via a finite-range
($r\leq{R}$) potential
\begin{equation}
V(r)\, =\, \frac{Z_1\, Z_2\, e^{2}}{r}\left(1 - \frac{r}{R}\right).
\end{equation}
Solving the classical problem \cite{Nagy94} for the collision time
($T$) we get
\begin{equation}
T(b,v,R)=\frac{2}{v}\left[\frac{\sqrt{R^2-b^2}}{c}+
\frac{Z_1Z_2e^2}{2E_1}\frac{1}{c^{3/2}}\ln\frac{\sqrt{c}\sqrt{R^2-b^2}+R+(Z_1Z_2e^2/2E_1)}{\sqrt{(Z_1Z_2e^2/2E_1)^2+c\,
b^2}}\right]
\end{equation}
where $E_1=M_1v^{2}/2$ is the energy of a heavy charged ($Z_1e$)
projectile moving with velocity $v$ and colliding at impact
parameter $b$ with a fixed center of {\it effective} charge $Z_2e$.
The dimensionless parameter introduced is $c=[1+(Z_1Z_2e^2/R)/E_1]$,
i.e., it is related to the ratio of potential and kinetic energies.
A closer inspection of this compact (and reasonable at short-range
solid-state conditions) form shows that practically one may use
\begin{equation}
T(b,v,R)\, \simeq{\alpha(v)\, \frac{R}{v}\,
\frac{E_1}{E_1+Z_1Z_2e^2/R}}\, \sqrt{1-\frac{b^2}{R^2}}
\end{equation}
to get a very acceptable estimation as a function of the heavy
projectile velocity $v$, with which $\alpha(v)$ is {\it decreasing}
from 4 (at $v\rightarrow{0}$) to 2 (at $v\rightarrow{\infty}$). If
we take $\beta(v)\simeq{T^{-1}(v)}$ and consider the energy change
determined above, we get a $\Delta
E(v)\sim{[\beta(v)]^{-2}}\sim{[\alpha(v)v]^2}$ character at very low
velocities. For the opposite, high-velocity, limit $\Delta
E(v)\sim{[\beta(v)]^{-2}}\sim{[\alpha(v)v]^{-2}}$ is the scaling,
which also looks quite reasonable physically. However, the detailed
work on the exactly determined non-linear energy changes is left for
a dedicated publication.

There, and also in Bohr's pioneering modeling
\cite{Bohr13,Merzbacher74} with time-dependent dipole fields of a
passing charge, the time-scale derived above using collision theory,
could play a more quantitative role to understand fine details
behind experimental predictions \cite{Moller04,Markin09} on energy
losses of slow ions in insulators. Furthermore, in the renewed field
of time-dependent energy losses, a proper time-scale also could help
to analyze further a remarkable theoretical prediction
\cite{Correa12} on the strong interplay of nuclear and electronic
stopping components of the observable total energy transfer. Indeed,
few-electron shake-up processes can be dominant at close encounters
in condensed matter. In such dynamical cases, one-electron
approximations, with double-occupancy for an auxiliary spatial
orbital, could result in inaccuracies.


%
\begin{acknowledgments}
We thank Professor P. M. Echenique for the very warm hospitality at
the DIPC. One of us (IN) is grateful to Professor P. Bauer and
Professor D. S\'anchez-Portal for useful discussions on energy
transfer processes in insulators. This work was supported in part by
the Spanish Ministry of Economy and Competitiveness MINECO (Project
No. FIS2013-48286-C2-1-P).
\end{acknowledgments}
%


\end{document}